# Leveraging VMware vCloud Director Virtual Applications (vApps) for Operational Expense (OpEx) Efficiency

Ramon Alvarez

Information Technology
Georgia Southern University
Statesboro, GA USA

Dr. Timur Mirzoev

Information Technology
Georgia Southern University
Statesboro, GA USA

Abstract— Virtualization technology has provided many benefits to organizations, but it cannot provide automation. This causes operational expenditure (OpEx) inefficiencies, which are solved by cloud computing (vCloud Director vApps). Organizations have adopted virtualization technology to reduce IT costs and meet business needs. In addition to improved CapEx efficiency, virtualization has enabled organizations to respond to business needs faster. While virtualization has dramatically optimized core IT infrastructures, organizations struggle to reduce OpEx costs. Because virtualization only addresses server consolidation, administrators are faced with the manual and resource-intensive day-to-day tasks of managing the rest of the data center – networking, storage, user management. This manuscript presents details on how leverage vApps based on a virtualized platform to improve CapEx efficiency in today's data center. The combination of virtualization and cloud computing can transform the data center into a dynamic, scalable, and agile resource capable of achieving significant CapEx and OpEx cost savings.

## I. INTRODUCTION

Before virtualization, organizations ran a single operating system and a single application on a physical x86 server, leaving most physical servers underutilized [1]. Virtualization technology addresses physical server underutilization by allowing to multiple virtual machines to run on a single physical machine side-by-side without compromising server performance [1]. Today, companies of all sizes have embraced virtualization to capitalize on physical server underutilization in an effort to reduce costs and remain competitive. Nearly 45 percent of servers within the organization are virtualized, and this number is expected to exceed 70 percent by 2015 [2]. There is no question that virtualization has been a major advancement in the IT industry by allowing organizations to significantly reducing IT costs while boosting efficiency and performance [3]. However, virtualization presents some limitations for organizations that need to react to fast changing business demands and environments [4]. For instance, as multitier applications require more complex resources and topologies, it is time-consuming and inefficient to manage a large of number of virtual machines (VMs) across the virtualized platform. Fortunately, cloud computing addresses these inefficiencies and provides organizations with new ways to implement traditional IT strategies [5]. Agility, scalability and innovation are the leading factors driving cloud adoption for many businesses [2].

Often times virtualization and cloud computing terms are used interchangeably, but it is important to point out that virtualization is not to be confused with cloud computing. Virtualization is the foundation for cloud computing [6]. While virtualization and cloud computing are highly coupled, there are significant differences between the two technologies. On one hand, virtualization refers to the mechanism of partitioning one physical server into multiple virtual servers in order to maximize server hardware. Cost savings are realized by consolidating multiple physical servers into virtual server instances that run on a single physical server [2]. On the other hand, cloud computing technology is based on virtualization to abstract virtualized resources and create large pools of resources comprised of compute, storage and network, which can be consumed from a self-service portal via the Internet [7].

Virtualization and cloud computing are powerful technologies that require advanced physical topologies in enterprise environments. For instance, a typical datacenter consists of multiple x86 virtualization servers, storage networks and arrays, IP networks, a management server, and desktop clients. Likewise, cloud computing platforms rely on one or more software-defined datacenters to provision ready-to-use services including VMs, storage, networking and security [8].

The need for such a complex network environment makes it difficult to build and deploy virtualized infrastructure to support multi-tier applications such as SAP ERP and robust e-commerce applications.

Cloud computing operates in one of several service models including Software as a Service (SaaS), Platform as a Service (PaaS) and Infrastructure as a Service (IaaS) [9]. In the SaaS service model, the service provider hosts applications, which are made available to customers over the Internet [10]. PaaS is a way for customers to lease hardware, storage and network resources as well as operating systems and applications via the Internet [10]. Unlike PaaS, IaaS is a service model in which customers rent only the IT infrastructure to support the organization's operations including servers, storage and networking resources [9]. Additionally, cloud computing technology can be deployed in a number of different ways including private, public, and hybrid clouds [9]. Private cloud refers to the provisioning of cloud resources for only a single organization, whereas public cloud is provisioned for open use by the general public [9]. Hybrid cloud is comprised of a combination of private and public clouds.

This manuscript analyzes commercial hardware and software for provisioning and deploying IaaS to support a multitier application scenario. IaaS deployment times will be compared in both virtualized and cloud environments to determine which approach is more efficient. Several techniques and approaches including catalog-based and template-based resource deployments will be discussed. There are several options and best practices for configuring multitier applications. However, the purpose of this study was only concerned with the provisioning and deploying virtualized infrastructure to support a multitier application and not with its installation and configuration. The multitier application scenario tested was comprised of three VMs, three virtual networks, and shared virtualized iSCSI.

### A. Statement of the Problem

While virtualization has provided many benefits, organizations are finding that virtualization alone falls short when provisioning and deploying virtualized infrastructure for complex multitier applications [3]. It is important for organizations to not only reduce IT costs through virtualization, but also to be able to leverage the existing virtualization infrastructure to support complex multitier applications. The cloud computing model offers organizations a way to maximize cost savings combined with increased IT agility [12]. Organizations can leverage existing virtualization infrastructure to build private clouds to deliver cloud service provider economics at scale, application provisioning in minutes, and automated operations management [3]. The operational expense savings results from improved manageability and the ability to use automation to work with large numbers of VMs with minimal manual effort.

### B. Research Limitations

Because cloud computing is a broad and complex topic, the scope of the research described in this manuscript is limited in the area of cloud computing. The research only focuses on the IaaS service model of cloud computing for the provisioning and deployment of infrastructure to support multi-tier applications. Other cloud computing service models including SaaS and PaaS are not discussed in this manuscript. Additionally, only the private cloud deployment model of cloud computing is analyzed in the study presented in this manuscript.

## II. LITERATURE REVIEW

As the number of workloads in a virtualized environment increase and the requirements for multitier become more complex, virtualization lacks the tools and technology to support workload scalability as well as complex multitier applications. Although it may seem as a relatively new technology, the concept of virtualization first appeared in the early 1960s. Virtualization technology is largely based on the concept time-sharing hardware developed by MIT in 1962 in partnership with IBM [11]. The prime issue that virtualization came to solve was the efficient utilization of the server hardware. A piece of software called hypervisor makes virtualization possible. The hypervisor decouples operation system and application from the server hardware allowing for multiple VMs to run simultaneously on the same physical server without negatively impacting performance [6]. As virtualization technology has become common in recent years, a large number of companies have entered the virtualization market. Although there are several smaller organizations which tend to specialize and develop their own take on virtualization technology, three manufacturers including VMware, Citrix and Microsoft have established leadership positions in the industry.

### A. VMware

VMware launched its first server hypervisor in 2001 [12]. VMware ESXi is the virtualization layer, or hypervisor, that runs on physical servers to abstract processor, memory, storage, and resources into multiple virtual machines [1]. As virtualization technology matured, it became possible to virtualize farms of physical servers.

VMware vSphere is a feature-rich suite of technologies that includes the VMware ESXi hypervisor at its core. VMware vSphere virtualizes and aggregates the underlying physical hardware resources across multiple systems and provides pools of virtual resources to the datacenter [10]. This allows organizations to build an entire virtual infrastructure with VMware vSphere, scaling across a large number of interconnected physical servers and storage devices [6].

A feature of the vSphere suite called vSphere Distributed Resource Scheduler (DRS) dynamically allocates and balances computing capacity across collections of hardware resources for virtual machines [14]. Like in DRS, vSphere Storage DRS allocates and balances storage capacity and I/O dynamically across collections of datastores [6]. DRS features represent a new way to manage hardware resources in the datacenter. Organizations no longer have to statically assign servers, storage, or network bandwidth resources to each application in the enterprise.

Another feature important feature is vSphere High Availability (HA), which provides high availability for VMs. If a server fails, affected virtual machines are restarted on other available servers that have spare capacity [1].

One of the most visible features of the vSphere suite is vMotion. In vMotion, powered-on VMs are migrated from one physical server to another with zero down time while preserving the virtual machine integration and business continuity [6]. When a VM is migrated using vMotion, only the VM's CPU and memory configuration files are moved; disk files remain in the original datastore [1].

If the goal is only to migrate the VM's disk files from one datastore to another, vSphere Storage vMotion allows for such a task. Migration with Storage vMotion moves the virtual disks or configuration file of a VM to a new datastore while the virtual machine is running [6].

Managing one or several hypervisor servers is not necessarily a daunting task. However, when the virtualized environment scales up and it incorporates hundreds of hypervisors, there is a need for centralized management. In order to provide a management layer for the virtualized infrastructure, VMware employs vCenter Server.

VMware vCenter Server is the central point for configuring, provisioning, and managing virtualized environments [1]. One of the core services provided by vCenter Server in a virtualized environment is the provisioning of VMs. Provisioning a VM consists of allocating CPU, memory, storage and networking resources out of a pool of virtualized resources across several hypervisors [13]. Another core service related to virtual machine provisioning is the concept of a virtual application or vApp. A vApp is a logical container used to group multiple VMs [13]. Because a vApp is a logical boundary for virtual machines, the vApp object can be managed as a single and separate entity in a virtualized environment. For example, a vApp can be used to package and manage a multitier application comprised of one or more front-end web servers, an application server, and a back-end database server. Management tasks such as powering on and off are applicable to the vApp object in addition to more advanced tasks such as cloning, or copying the vApp.

*B. Citrix*

Citrix has been a leader in the application virtualization market since the early 1990s, but it entered the server virtualization market in 2009 with its XenServer offering [15]. It is important to note that Xen originated as a research project at the University of Cambridge let by Ian Pratt, who later founded XenSource, Inc. [15]. Citrix later acquired XenSource, Inc. and XenServer emerged as an open-source virtualization technology.

Unlike VMware and Microsoft, the XenServer technology is developed and maintained by the community as free software, licensed under the GNU General Public License (GPLv2) [15]. XenServer is the hypervisor layer that runs directly on the physical server hardware. The first guest operating system called *Controller Domain* or *dom0* is a secure, privileged VM that runs the XenServer management tool stack known as xapi [16] Guest VMs created and running on XenServer gain access to devices on the underlying physical server via the *Control Domain* [17].

Citrix's virtualization offering includes several technologies such as XenMotion, XenCenter Management, XenServer Conversion Manager and other critical data center automation tasks. The Citrix XenMotion feature allows organizations to migrate powered-on VMs from one physical server to another with zero down time and without disruption the operation system or applications [17]. The ability to shift workloads on demand makes it possible for organizations to rebalance workloads within the virtualized environment or to perform physical server maintenance without affecting service levels.

The XenCenter Management feature provides a centralized point of management for all VM monitoring, management and general administration functions [17]. Unlike other technologies such as VMware vCenter Server, by design XenCenter is a highly available management architecture because it all management and configuration data across all servers without the need for a separate database [17].

In the absence of a standardized virtualization platform, manufacturers develop their own technologies and protocols to support their virtualization technologies. In the case where an organization switches virtualization technologies vendors, there is a need to migrate the existing virtual machines from one vendor's virtualization platform to another. XenCenter Conversion Manager offers a way for organizations to convert VMware VMs into XenServer VMs [17]. XenServer offers protection for VMs via high availability. VMs are automatically restarted not only if the physical server fails, but also if the VM itself or the hypervisor fail [17].

One way XenServer helps organizations leverage existing and new server equipment is by creating heterogeneous resource pools. New servers with relatively newer CPU configurations are able to join existing resource pools containing servers with slightly older CPU configurations [17]. CPU features are configured to appear as providing a different make, model, or functionality than what the features actually are as long as the CPU is from the same vendor [18].

*C. Microsoft*

Microsoft had a small presence in the virtualization market since 2003 by offering products such as Virtual PC and Virtual Server [19]. However, it was not until 2008 that Microsoft released its Hyper-V as its enterprise-level virtualization offering [19].

Microsoft has taken a different approach to virtualization with Hyper-V. Hyper-V is based on the *microkernelized* hypervisor, which means it runs on top of the Windows Server operating system [20]. Using this approach, a host operating system, or *parent partition*, communicates directly with the physical hardware to provide management functionalities and the drivers for the hardware [20]. As VMs are created in Hyper-V, the parent partition creates *child partitions* to host guest operating systems [21]. The *parent partition* is responsible for managing and handling all VM requests to access physical hardware.

One advantage of Hyper-V is that it can leverage existing Microsoft technologies. For instance, Hyper-V fault tolerance is based on the Windows Server Failover Clustering feature to increase the availability of virtual machines and applications. A failover cluster is a group of independent computers that work together as a single entity [21]. Another feature called Live Migration can move running VMs between Hyper-V server hosts without disruption of service or perceived downtime [21]. Live Migration makes it possible to keep VMs online even during physical server maintenance periods.

Hyper-V Replica is a feature that can asynchronously replicate a VM running from one location to another, including across the WAN [19]. If the primary site experiences either a planned or unplanned outage, the replicated VM in the replica site is brought online to maintain service and business continuity.

Storage Quality of Service (QoS) in Hyper-V is the ability to limit individual virtual machines to a specific level of I/O throughput. Hyper-V Network Virtualization is another feature that addresses the limitations and complexities seen in typical environments using virtual LANs (VLANs) [19]. For instance, Hyper-V Network Virtualization allows for VMs to be migrated between Hyper-V hosts on a different network subnet. In addition, Hyper-V Network Virtualization allows organizations to re-locate VMs without having to change the existing IP addressing scheme of the VMs [20]. Multi-tenancy with overlapping IP address ranges is possible with Hyper-V Virtualization to isolate VMs at the network layer even when the VMs share the same IP address scheme [21].

To manage the virtualized environment, Microsoft leverages System Center Virtual Machine Manager (SCVMM). SCVMM is designed for centralized management of large number of Hyper-V host servers, network and storage resources [21].

### D. Cloud Computing

Today, many organizations have implemented virtualization to reduce cost and increase the efficiency of their existing IT infrastructure. VMware, Citrix, and Microsoft are the leading manufacturers in the virtualization industry providing different approaches to virtualization. However, one constant remains true; virtualization is the catalyst for cloud computing [19]. Similar to the concept of virtualization, the concept of cloud computing dates back to the 1950s and 1960s. In the 1950s, the concept of cloud computing can be traced back to large-scale mainframes used at schools and corporations. Multiple users were able to access the mainframe via "dumb terminals" [7]. For instance, in the 1960s J.C.R. Licklider described his vision for the future of computing where everyone on the globe would be interconnected and access programs and data at any site, from anywhere [22]. In 1961, John McCarthy made a prediction stating that "*Computation may someday be organized as a public utility*", which fits to the contemporary working definition of cloud computing [23]. Because mainframes were expensive pieces of equipment, it was impossible to afford a mainframe for each employee or scholar. Instead, the organization shared mainframe access to get a better return on investment.

As the costs of server hardware slowly came down, more users were able to purchase their own dedicated servers. Over time, a new problem emerged; one server is not enough to provide the resources needed. This caused a shift in the way servers were used. Servers shifted from being expensive and shared to being more affordable and aggregated. By installing and configuring a piece of software called a hypervisor across multiple physical servers, a virtualized environment can present all of aggregated resources as a single physical node. Cloud computing helps to visualize such environment since the sum of the parts seemed to become a large pool of computing resources which could then be segmented out as needed [5]. Because cloud computing depends on a virtualized environment, extending the computing capacity becomes a matter of adding physical servers to the existing virtualized environment. Once properly configured, the additional resources become part of the bigger system, or the cloud environment.

### E. Scripting

As virtualized environments grow in size and complexity, the task of managing and administering resources becomes daunting. Scripting has been a solution used to facilitate the automation of repetitive and manual tasks in the virtualized environment. VMware, for example, implemented PowerCLI as a scripting mechanism [24]. Some of the tasks that administrators can automate with PowerCLI include automating of vCenter Server as well as hypervisor ESXi deployment and configuration; automating storage and networking; and creating virtual machines. However, there are inherent problems with using scripting to automate the virtualized environment. Administrators must possess a deep understanding of the scripting language and constructs before they can be effective using it. Also, the risk exists for accidental configuration changes made to the virtualized environment via scripting, which could have a negative impact on the environment. Also, depending on the task at hand, the script can vary in length from a couple of lines to potentially thousands of lines.

### III. METHODOLOGY

### A. Research Design

In order to compare Infrastructure as a Service (IaaS) deployment times in a virtualized environment and in a cloud environment, two scenarios were created using VMware products. The author of this manuscript used VMware products including VMware vSphere 5.1 and VMware vCloud Director 5.1 as the methodology instruments because he had readily access to them. The experiment was conducted Georgia Southern University using the Cloud Computing Research Laboratory (http://cloud.georgiasouthern.us). A base workload comprised of three VMs, three networks, and shared virtualized iSCSI storage was used in both scenarios as shown in Table 1.

TABLE 1. BASE WORKLOAD USED FOR IaaS PROVISIONING AND DEPLOYMENT

| Hostname | OS | Role | vCPU | RAM (GB) | Disk Size (GB) | Eth0 | Eth1 | Eth2 |
|----------|-----|------|------|----------|----------------|------|------|------|
| ACME-VM1 | Windows 2008 R2 64-bit | Front-end web server | 1 | 4 | 60 | 192.168.0.10 | 172.16.30.10 | 10.1.1.10 |
| ACME-VM2 | Windows 2008 R2 64-bit | Application server | 1 | 4 | 100 | 192.168.0.20 | 172.16.30.20 | 10.1.1.20 |
| ACME-VM3 | Red Hat Enterprise Linux 6.4 64-bit | Back-end database server | 1 | 2 | 40 | 192.168.0.30 | 172.16.30.30 | 10.1.1.30 |
| OpenFiler | OpenFiler 2.99 64-bit | NAS/SAN appliance | 1 | 4 | 200 | 192.168.0.40 | 172.16.30.40 | 10.1.1.40 |

## B. Techniques of Data Gathering

The following resources where used during this experiment:

1. Two VMware ESXi hypervisor servers (version 5.1.0 Build 1065491) based on Dell PowerEdge R610 with the following specifications for each server:

   ➢ 8 Intel Xeon E5520 2.26GHz CPUs

   ➢ 98 Gigabytes DDR Memory

   ➢ 12 Gigabit Network Interfaces

2. One VMware vCenter Server (version 5.1.0 Build 1064983)

3. One vSphere Client (version 5.1.0 Build 1064113)

4. VMware vCloud Director (version 5.1.1.868405)

5. Firefox web browser (version 25.0)

6. Microsoft Excel 2010

## C. Experiment Assumptions

1. A vApp object will be used to group the workload in both virtualized and cloud environments. The vApp will be configured with its default settings, and it will be assumed to be functional.

2. The underlying physical networking infrastructure will be assumed to be fully functional, and a distributed virtual switch (dvSwitch) will be available in both scenarios.

3. The cloud provider vDC in VMware vCloud Director 5.1 will be assumed to be functional configured with default settings. Also, it will be connected to the same vCenter Server.

4. Once VMs are provisioned in both scenarios, the task of installation of the operating system will not be counted as part of the IaaS deployment time.

5. Licensing of VMware products will not be a limitation for this experiment.

## D. IaaS Provisioning and Deployment

The IaaS provisioning and deployment process is divided in several sections. Each section represents a series of tasks to be completed. Completion of these validates the provisioning of virtualized infrastructure in this experiment for a multitier application as shown in Figure 1.

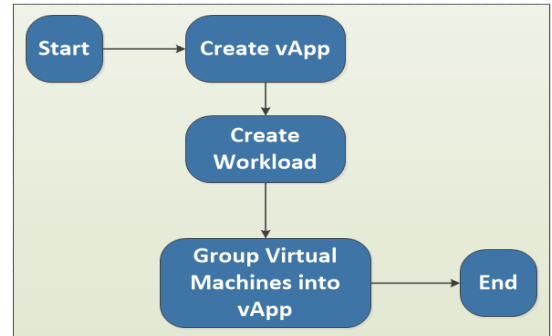

Figure 1. IaaS provisioning and deployment process.

It is important to note that product solutions provided by VMware are offered using a licensing model in order to generate revenue. For the purpose of the experiment conducted in this research, licensing was not a relevant component. However, in the real world licensing is an important consideration because it can have an impact on capacity planning and advanced feature availability.

## E. Virtualized Environment

Each step in the IaaS provisioning and deployment process will be timed using a stopwatch. At the end of each step, the time will be recorded in a Microsoft Excel 2010 spreadsheet. A vApp object will be created in vSphere 5.1 using default settings. Then, each of the four VMs included in the test workload will be provisioned. After all VMs have been provisioned, they will be grouped into the previously created vApp. Lastly, each of the three networks will be provisioned and each of the VMs will be connected to each virtual network.

## F. Cloud Environment

The steps for provisioning and deployment IaaS in the cloud environment are similar to the steps in the virtualized environment with the exception of a couple of deviations. Each step will be timed using a stopwatch. At the end of each step, the time will be recorded in a Microsoft Excel 2010 spreadsheet. A vApp object will be created in vCloud 5.1 using default settings. Then, each of the four VMs included in the test workload will be provisioned inside the vApp. Unlike in vSphere, a vCloud VM cannot exist outside a vApp. Lastly, each of the three required networks will be provisioned and each of the VMs will be connected to each virtual network.

## IV.   FINDINGS

The recorded times during the process of provisioning virtualized infrastructure for the test workload as shown in Table 2.

TABLE 2. IAAS PROVISIONING AND DEPLOYMENT ACTUAL TIMES.

|          | vApp   | VMs        | Net 1       | Net 2       | Net 3       |
|----------|--------|------------|-------------|-------------|-------------|
| vSphere  | 13 sec | 4 min 45 sec | 1 min 29 sec | 1 min 07 sec | 1 min 12 sec |
| vCloud   | 30 sec | 2 min 41 sec | 2 min 16 sec | 2 min 40 sec | 2 min 02 sec |

Some important capabilities exist in VMware vCloud Director that do not exist in VMware vSphere. For instance, once created and populated with VMs, a vCloud vApp can be stored in a template and saved into a catalog for future provision. When the vApp is deployed from the catalog, not only will the group of VMs be included but also the configured networks. A vApp template enables users in an organization to quickly deploy entire vApps that have already been configured. By creating a collection of vApp templates, users can avoid the time required to set up and configure an environment for use. It also can enable the cloud administrator to define standardized versions of the vApps the organization wants users to deploy. Although a vApp can be cloned in vSphere, the networking resources will have to be configured manually since they are not part of the vApp.

Another advantage of VMware vCloud over vSphere is the self-service capability. Users login to the cloud environment from a browser to provision infrastructure tasks assuming they have the appropriate levels of privileges. In vSphere, only the administrator of the virtualized can execute the provisioning of virtualized infrastructure. Although it would be possible to configure varying levels of user access to the virtualized environment, this approach would not be ideal when dealing with a large number of users. Adding to the administrative burden in vSphere, if access to the virtualized environment were to be provided via a web browser, this would mean that access would be limited to internal access. In the unlikely event the organization opted to provide external access to the vSphere environment, the administrative burden would increase since additional changes including firewall rules and domain name resolution changes would have to be made in order to provide external access. Often times, this is a security risk that most organizations would not take.

In addition, VMware vCloud Director provides a mechanism to allocate virtualized resources. These resource allocation models help organizations control the quality of service and the costs associated with the virtualized resources. For instance, when creating a cloud organization resources can be assigned in one of three available allocation models. One allocation model is Allocation Pool in which only a percentage of the resources allocated is committed to the organization virtual datacenter (vDC). Because the organization has control over the percentage value, it allows for the overcommitting of resources across multiple virtual providers for different organizations. Another allocation model is Pay-As-You-Go, which only commits resources when users create vApps in the organization. The organization can specify a percentage of resources to be guaranteed, which allows for overcommitting resources. The Reservation Pool model is the least elastic of the three because allocated resources are committed immediately. In addition, provider vDCs can be merged in vCloud Director, which provides a great level of resource flexibility. Merging of resources in this fashion is not possible in vSphere.

Another capability found in vCloud Director that does not exist in vSphere is the ability to configure network settings for the vApp in fenced mode. The term fencing in vCloud enables multiple virtual machines in various vApps to be powered on without conflict by isolating the IP and MAC addresses of the virtual machines.

Solid understanding of virtualized concepts and technology is required in order to advance to the level of cloud computing. This is important because virtualization changes the way traditional IT is approached. In addition, commercial training can be expensive, and if the organization employs a large group of cloud administrators, it could represent a significant expense. Depending on the feature set required, the cost of purchasing the appropriate licensing for VMware vCloud Director software can be substantial.

## V.   CONCLUSIONS

Over the past decade, virtualization has become a vital pillar of the IT infrastructure for many organizations worldwide. Not only has virtualization been able to provide significant savings in CapEx, but also it has allowed organizations to achieve OpEx savings. However, despite the many benefits provided by virtualization, organizations are discovering that virtualization alone may not be enough to support the today's increasingly complex application environments. Multitier applications are no longer constrained to a single server, but rather the application is distributed across multiple servers that together work as a single unit. In addition, the underlying complexity of the environment topology for multitier applications requires an increasing number of interconnected networks in order to provide the necessary functionality. The research in this manuscript was conducted to determine if a virtualized environment or a cloud environment would be more efficient in providing virtualized infrastructure to support a multitier application. The research was conducted using the Cloud Computing Research Laboratory at Georgia Southern University.

Based on the findings of this study, it is apparent that as the complexity of the application environment increases, virtualized infrastructure delivered through cloud computing is more efficient than through a virtualized environment. An example of increased complexity could consist of a multitier application with five or more networks interconnecting storage and the various application components such as front-end and back-end servers as well as database servers. In a virtualized environment such as VMware vSphere, such networks must be built from the ground up as part of the infrastructure provisioning process. In contrast, in a private cloud environment such as VMware vCloud Director (vCD), the virtualized network resources are further abstracted to include networking as part of the provisioning process. For instance, a cloud vApp can be created with as many networks as needed as long as the network pool has enough resources. It would be hard to replicate this behavior in a vSphere vApp as networks are constrained by the underlying physical network topology.

By extending the existing virtualization platform, vCD provides features that are not available in vSphere. For instance, vCD introduces the catalog construct, which is storage repository for vApps and media. Catalogs are available to users within the organization and can be shared with other organizations within the cloud environment. Users can quickly provision vApps storage in a catalog in a matter of minutes including the networking topology regardless of its complexity.

Another important aspect of vCD is the self-service portal. Unlike users in a vSphere environment who depended on the administrator, users in vCD are empowered to deploy preconfigured services or build a complete application stack with a few clicks via the self-service portal. In addition, resource provisioning is approached differently in vCD than in vSphere. Resources are allocated in one of three available models including Allocation Pool, Pay-As-You-Go, and Reservation Pool. In vSphere, resource allocation models do not exist. The automation behind the pooling of resources in vCD allows the efficient delivery of resources to organizations when requested and the efficient retrieval of resources when resources are no longer needed.

However, it is important to point out that despite the many benefits, vCD also presents some challenges. The principal challenge is that vCD can only manage a virtualized environment that is run on VMware vSphere. This presents a significant problem for companies that are not necessarily using VMware technologies. It is important to consider such a limitation when architecting a virtualized and cloud strategy. Ideally, companies need a private cloud solution that can leverage the existing virtualization platform regardless of vendor. A vendor neutral capable of connecting users with the resources provided by the virtualized platform would certainly increase the appeal of cloud computing adoption. Another challenge for companies adopting vCD is trained personnel. vCD administrators must possess a deep understanding of vSphere environments as well as vCD concepts such as virtual datacenters, external networks, routed networks, fenced networks among other important constructs. For instance, if a vCD administrator maps a provider virtual datacenter (vDC) to a vSphere resource pool instead of a vSphere cluster, the provider vDC may experience problems if the same resource pool is used by another provider vDC or by an Org vDC.

In conclusion, cloud computing technologies such as VMware vCloud Director extend existing vSphere virtualization environments to create new levels of computing resource abstraction. Cloud vApps are a way to encapsulate higher levels of computing architecture and resource complexity, which can then be deployed in a matter of minutes with less effort. Cloud computing opens a wide range of opportunities for organizations to think about IT solutions and possibilities. However, it is important to extend the testing performed in this manuscript to other virtualization and private cloud computing solutions such as Citrix XenServer and Microsoft Hyper-V. It is necessary to consider how other virtualization-private cloud technologies compare to VMware's approach when provisioning virtualized infrastructure to support multi-tier applications. Although VMware is a leader in virtualization technology, IT infrastructures across enterprises are often comprised of a hybrid mixture of technologies or non-VMware technology all together. As cloud computing technology matures, there may be significant improvements in the way virtualized infrastructure is provisioned and managed to support complex environments.

## AUTHORS PROFILE

Ramon Alvarez is a graduate student in the Masters of Science in Applied Engineering with a concentration in Information Technology. In addition, he is a research assistant to Dr.Timur Mirzoev, Assistant Professor in Information Technology at Georgia Southern University. Mr. Alvarez is the prime cloud administrator for the Cloud Computing Research Laboratory (CCRL) at Georgia Southern.